\documentstyle[aps,prd,epsfig,floats]{revtex} 

\draft

\begin{document}
\twocolumn[\hsize\textwidth\columnwidth\hsize\csname
@twocolumnfalse\endcsname
\title{Helioseismological constraint on solar axion emission}

\author{H.~Schlattl and A.~Weiss}
\address{Max-Planck-Institut f\"ur Astrophysik,
Karl-Schwarzschild-Str.~1, 85740 Garching, Germany}

\author{G.~Raffelt}
\address{Max-Planck-Institut f\"ur Physik 
(Werner-Heisenberg-Institut),
F\"ohringer Ring 6, 80805 M\"unchen, Germany}

\date{\today}

\maketitle

\begin{abstract}
  Helioseismological sound-speed profiles severely constrain possible
  deviations from standard solar models, allowing us to derive new
  limits on anomalous solar energy losses by the Primakoff emission of
  axions.  For an axion-photon coupling
  $g_{a\gamma}\alt5\times10^{-10}~{\rm GeV}^{-1}$ the solar model is
  almost indistinguishable from the standard case, while
  $g_{a\gamma}\agt10\times10^{-10}~{\rm GeV}^{-1}$ is probably
  excluded, corresponding to an axion luminosity of about
  $0.20\,L_\odot$.  This constraint on $g_{a\gamma}$ is much weaker
  than the well-known globular-cluster limit, but about a factor of 3
  more restrictive than previous solar limits. Our result is primarily
  of interest to the large number of current or proposed search
  experiments for solar axions because our limit defines the maximum
  $g_{a\gamma}$ for which it is self-consistent to use a standard
  solar model to calculate the axion luminosity.
\end{abstract}
\vskip2.0pc]


\section{Introduction}

Stars are powerful sources for low-mass weakly interacting particles.
The backreaction of this energy-loss channel on the properties and
evolution of stars has been extensively used to constrain nonstandard
neutrino couplings and the properties of axions and other hypothetical
particles~\cite{RaffeltBook}. While the Sun no doubt is the star best
known to us, it does not provide the most restrictive limits on new
particle properties. Still, it remains of interest what the Sun can
tell us about nonstandard particle-physics assumptions.

Much has improved in our knowledge of the Sun over the past decade
because its interior properties have become accessible by
helioseismological methods and by a host of solar neutrino
experiments.  It used to be that a solar model was fixed by adjusting
its presolar helium abundance and the mixing-length parameter such as
to reproduce the observed luminosity and radius at an age of about
$4.6\times10^9~{\rm years}$. In this way even rather dramatic
modifications could be accommodated. For example, a nonstandard
energy-loss channel in the form of axion emission could have been
essentially as large as the solar photon luminosity so that the
present-day Sun could have been close to its main-sequence
turn-off~\cite{RaffeltDearborn87}.  Naturally, one expects that the
precise helioseismological sound-speed profiles which have become
available over the past few years will provide far more severe
restrictions on possible present-day solar models.

\begin{figure}[h]
\hbox to\hsize{\hss\epsfxsize=3.5cm\epsfbox{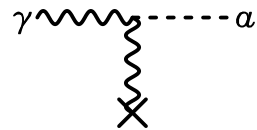}\hss}
\caption{Primakoff production of axions in the Sun.
\label{fig:primakoff}}
\end{figure}

As a specific example for a nonstandard energy-loss channel we
consider the Primakoff conversion of axions in the Coulomb fields of
charged particles, $\gamma+Ze\to Ze+a$ (Fig.~\ref{fig:primakoff}).
This case is of particular interest because one can search for the
solar axion flux by the reverse process where an axion oscillates into
an X-ray in a long dipole magnet which has been oriented toward the
Sun, the so-called ``helioscope'' method~\cite{Sikivie,Bibber}. An
alternative approach uses the Primakoff backconversion in a germanium
crystal where one can achieve a large enhancement in some directions
in analogy to Bragg diffraction~\cite{Germanium}.  In either case it
is of interest to know the maximum axion-photon coupling strength
which is compatible with well-established solar properties. For
example, an early experimental search for solar axions~\cite{Lazarus}
was not sensitive enough to detect these particles even if the
coupling strength had been so large as to push the present-day Sun
close to its main-sequence turn-off.  Put another way, the limit found
in this early experiment was not self-consistent in that it required a
larger axion emission from the Sun than is compatible with its age.  A
surprisingly large number of current~\cite{Tokyo,Solax} and
proposed~\cite{Novosibirsk,LHC} solar axion search experiments has
recently emerged. This experimental activity motivates us to examine
solar axion limits in the light of helioseismological information.

In Sec.~II we describe our standard solar model and compare it with
recent helioseismological data. In Sec.~III we construct models of the
present-day Sun including axionic energy losses and derive a new solar
limit on the axion-photon coupling constant.  In Sec.~IV we discuss
our results in the light of other astrophysical limits and in the
context of solar axion search experiments.

\section{Solar and Seismic Model}

The solar model used in this paper was computed with the GArching
SOlar Model (GARSOM) code, which has been described in its numerical
and basic physical details elsewhere~\cite{Schlattl97}.  It was
shown~\cite{Turck98} that it agrees in its gross features with other
contemporary up-to-date standard solar models.  In particular, it uses
the latest OPAL opacities and equation of state~\cite{OPEOS,OPOPAC}
and takes into account particle diffusion of hydrogen, helium and a
number of heavier elements (e.g.\ C, N, O).  We usually follow
the evolution beginning from the pre-main sequence to an age
of 4.57Gyr (``best model''). To derive the differential effect of axion 
emission on solar models it is, however, sufficient to calculate
models including only H/He-diffusion starting from the zero-age main 
sequence (ZAMS).


The main difference to other comparable solar models is our treatment
of the atmosphere and the outermost superadiabatic convective layers
down to an optical depth of 1000, where the stratification of a
two-dimensional hydrodynamical simulation~\cite{Hydro} is used.  In
the deeper layers convection is treated according to
Ref.~\cite{Canuto} even though the temperature gradients are almost
perfectly adiabatic and thus independent of the convection theory.

Since the first publication of our solar model~\cite{Schlattl97} the
code has been revised. Apart from some minor details the
following numerical treatments have been changed:
particle diffusion and the nuclear network are now solved
simultaneously in the same system of equations instead of following a
sequence of burning-mixing-burning between two models of successive
age. The opacity interpolation (a two-dimensional bi-rational spline
with one free parameter) now resembles closely a standard cubic-spline
interpolation, because we found that the strong spline damping we
previously used to prevent unphysical oscillations of the
interpolation function degraded the agreement of our models with
seismic ones~\cite{Schlattl98}. Furthermore, the reaction rate for
$^3{\rm He}(^3{\rm He},2p)^4{\rm He}$~\cite{LUNA} and the solar radius
($R_\odot = 6.95508\times 10^{10}~{\rm cm}$~\cite{Brown98}) were
updated.

\begin{figure}[h]
\hbox to\hsize{\hss\epsfxsize=9cm\epsfbox{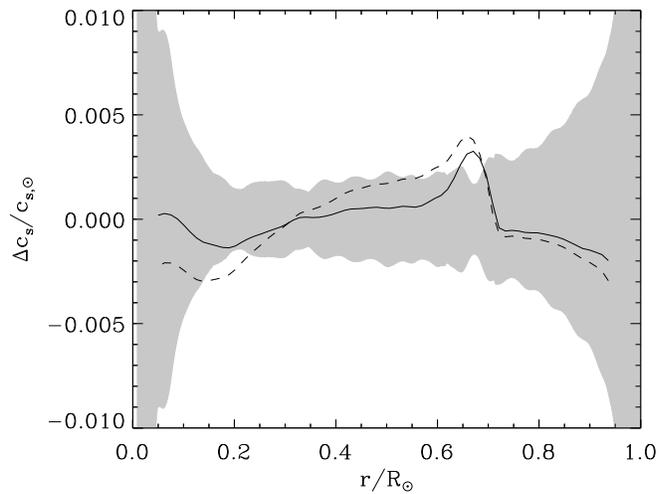}\hss}
\caption{Difference in sound-speed profiles of present-day solar
models compared to the seismic model. The shaded are
reflects the uncertainties in the infered sound speed of the
seismic model. The solid line shows our best solar model,
the dashed line the reference model ($g_{10}=0$) used for our analysis.
\label{fig:cprofile}}
\end{figure}

Theoretical solar models are compared with seismic ones which are
constructed by the so-called inversion method from measured p-mode
frequencies.  In Fig.~\ref{fig:cprofile} we show the difference in
sound speed of our best solar model (solid line) compared to the
seismic model from~\cite{Basu97}. The dashed line shows a solar model
which was calculated from the ZAMS including only H/He-diffusion ($g_{10}=0$).
The latter is used in the following as reference model to illustrate
the effect of axion emission. The shaded area shows the 
uncertainty of the derived profile of the sound speed ($c_{\rm s}$) according 
to Ref.~\cite{Scilla97}. They were quoting the errors in the
quadratic isothermal sound speed. To get the respective errors in
$c_{\rm s}$ we neglect the uncertainties from the adiabatic index 
$\Gamma_1$, as these contribute barely to the total error~\cite{Antia96}.
There are three sources for the total uncertainty: (i)~the errors of the 
measured frequencies; (ii)~the dependence of the final seismic model on the
starting model; and finally, (iii)~the uncertainties in the
regularization procedure of the inversion method.  The three
uncertainties were determined in a conservative way: each error
interval was doubled and then their absolute values were
added~\cite{Scilla97}. This was done because the parameter study may
not have been exhaustive. The measured frequencies being very precise,
the main uncertainties arise from the inversion method itself.
Previous studies by~\cite{Antia96} found
an uncertainty of approximately $5 \times 10^{-4}$ for
$0.2 < r/R_\odot < 0.8$ for each of the three error sources 
quoted above. Applying the same procedure for determining the conservative
errors gives a value of $3\times 10^{-3}$ slightly bigger than the 
uncertainties found by~\cite{Scilla97}. \cite{Antia98} remark that
seismic models do not really constrain solar model
for $r/R_\odot<0.05$.

The deviation of our standard solar model from the seismic model
by~\cite{Basu97} (solid
line in Fig.~\ref{fig:cprofile}) remains almost everywhere within the error
range. The deviation is
very similar to that of other comparable solar models in the
literature. In particular, we also find a large deviation immediately
below the convective zone which might be indicative for either a
slight error in the opacities or for overly effective helium diffusion
out of the convective envelope~\cite{Richard96}. 

In the first row of Table~\ref{tab:runs2} we list the expected
counting rates for the gallium and chlorine experiments as well as the
flux of $^8$B neutrinos at the (Super)Kamiokande detector.  Assuming
flavor conversion of two neutrino types as the solution of the solar
neutrino puzzle we show the parameter space in Fig.~\ref{fig:mixing}
for which the measured neutrino fluxes of all experiments are
reproduced. The upper panel shows the MSW-solutions~\cite{Mikey85},
the lower one the solution for vacuum oscillations (shaded areas). In
the calculation of the MSW-solutions we include the average
earth-regeneration effect~\cite{Earthreg}, and for the
vacuum-oscillations the annual Sun-Earth distance variation.  Our 95\%
confidence level regions agree with those in the literature.

We have verified by test calculations that including metal diffusion
and the new value for $S_{17}(0)$ of the $^7{\rm Be}(p,\gamma)^8{\rm
  B}$ reaction of $0.019~{\rm keV~barn}$~\cite{Adel98} instead of
$0.0224~{\rm keV~barn}$~\cite{NucRate} leads to almost the same
parameter regions in the $\Delta m^2$-$\sin^22\theta$-plane as for
the reference solar model.



\section{Solar Models with Axion Losses}

We now calculate self-consistent solar models where we include 
energy losses by axion emission from zero age to the present-day
Sun including only H/He-diffusion. Axions~\cite{Axions} generically
have a two-photon coupling
vertex in full analogy to neutral pions. The interaction Lagrangian
can be written in the form
\begin{equation}
{\cal L}_{a\gamma}=g_{a\gamma} {\bf B}\cdot{\bf E}\,a,
\end{equation}
where $g_{a\gamma}$ is a constant with the dimension (energy)$^{-1}$,
${\bf E}$ and ${\bf B}$ are the electric and magnetic fields, and $a$
is the axion field. In keeping with previous works
we will always use the dimensionless parameter
\begin{equation}
g_{10}\equiv g_{a\gamma}/10^{-10}~{\rm GeV}^{-1}
\end{equation}
to characterize the coupling strength. The energy-loss rate as 
a function of $g_{10}$ is discussed in Appendix~A. 

\begin{table}[b]
\caption{\label{tab:runs1}
Solar models with axion losses.}
\smallskip
\begin{tabular}[6]{cccccccc}
$g_{10}$ & $L_a/L_\odot$ & $Y$ & $Y_c$ & 
$\rho_c$ & $T_c$ & 
$|\Delta c_{\rm s}/c_{\rm s,0}|_{\rm max}$\\
&&&& 
$[{\rm g/cm^{3}}]$ &$[{\rm 10^7~K}]$&\\
\noalign{\vskip3pt\hrule\vskip3pt}
0   &  0    &  0.266 &0.633 & 153.8 & 1.563 & 0.00\%\\
4.5 &  0.04 &  0.265 &0.641 & 158.0 & 1.575 & 0.16\%\\
10  &  0.20 &  0.257 &0.679 & 177.5 & 1.626 & 0.81\%\\
15  &  0.53 &  0.245 &0.751 & 218.3 & 1.722 & 1.82\%\\
20  &  1.21 &  0.228 &0.914 & 324.2 & 1.931 & 3.12\%\\
\end{tabular}
\bigskip\bigskip
%
\caption{\label{tab:runs2}
Neutrino detection rates.}
\smallskip
\begin{tabular}[5]{cccc}
$g_{10}$ & Ga & Cl & $^8$B \\
& [SNU] & [SNU] & [$10^6~\rm s^{-1}cm^{-2}$]\\
\noalign{\vskip3pt\hrule\vskip3pt}
0   &   127   &   8.0   &   5.5  \\
4.5 &   136   &   9.3   &   6.6  \\
10  &   184   &   17.6  &   13.0 \\
15  &   323   &   48    &   37   \\
20  &   806   &   161   &   127  \\
\end{tabular}
\end{table}

\begin{figure}[h]
\hbox to\hsize{\hss\epsfxsize=9cm\epsfbox{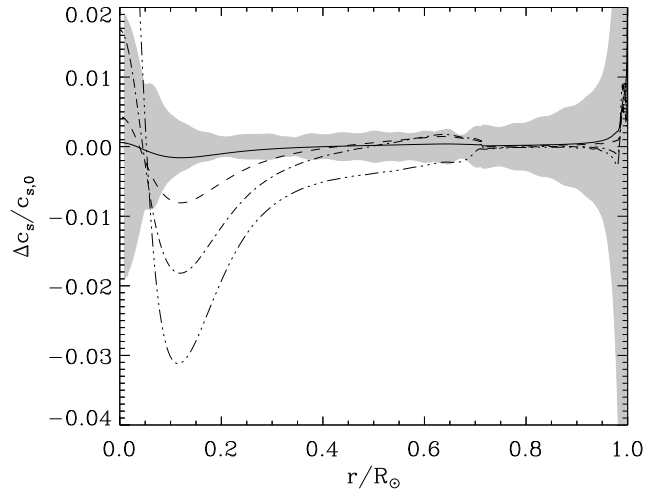}\hss}
\caption{Difference in sound-speed profiles of present-day solar
models with axion losses compared to the reference model in
the sense (Reference$-$Model)/Reference. 
Different line types correspond to different values of the
axion-photon coupling constant: $g_{10}$=4.5 (solid line), 10
(short-dashed), 15 (dash-dotted), 20 (dash-dot-dot-dotted).
The shaded area is the same as in Fig.~\ref{fig:cprofile}.
\label{fig:uprofile}}
\end{figure}

In Table~\ref{tab:runs1} we summarize the characteristics of the solar
models which include axion losses for several values of $g_{10}$. We
show the present-day axion luminosity $L_a$, the presolar helium
abundance $Y$, and the helium
abundance, density, and temperature at the solar center. In
Fig.~\ref{fig:uprofile} we show the deviation of the sound speed
profiles of the axionic solar models from a model without 
axion loss ($g_{10}=0)$.
This deviation shows a local maximum at around $r\approx
0.1\,R_\odot$, a region where both the models are most sensitive to
axion emission and the seismic models are well-constrained.  The
fractional deviation $|\Delta c_{\rm s}/c_{\rm s,0}|_{\rm max}$ at this
local maximum is listed in Table~\ref{tab:runs1}. 

\begin{figure}[t]
\hbox to\hsize{\hss\epsfxsize=9cm\epsfbox{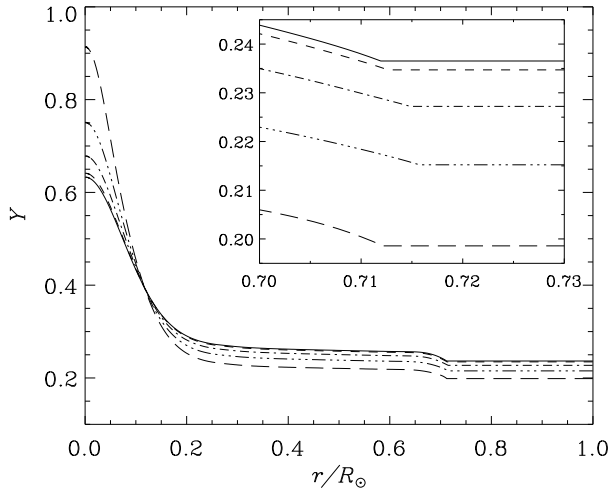}\hss}
\caption{Helium profiles of present-day solar models with
different axion losses. The insert shows a magnification of the
region around the bottom of the convective envelope. The line styles
correspond to those of Fig.~\ref{fig:cprofile}.
\label{fig:yprofile}}
\end{figure}

In Fig.~\ref{fig:yprofile} we show the helium profiles of our axionic
solar models.  The depth of the convective envelope is affected very
little by the axion loss---all models are within the value predicted 
from helioseismology of 0.710--0.716$\,R_\odot$~\cite{Joerg91}. On the
other hand, the present surface helium content depends on the initial
one, reduced by diffusion in the course of the evolution by about
0.03.  Axionic models with $g_{10} \geq 10$ have a surface helium
abundance significantly lower than 0.238, the smallest value allowed
from helioseismology~\cite{Scilla97}.

\begin{figure}[ht]
\hbox to\hsize{\hss\epsfxsize=9cm\epsfbox{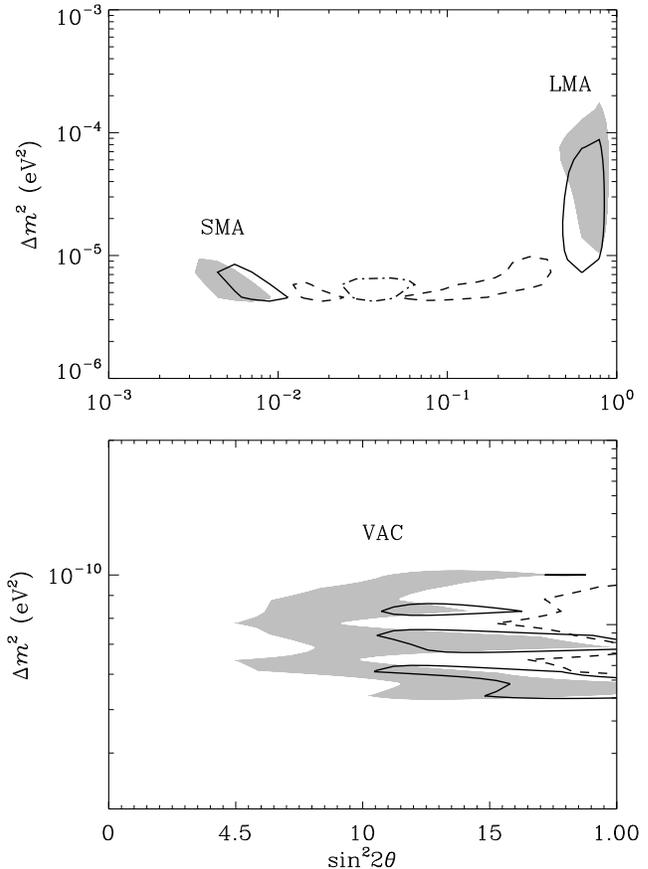}\hss}
\caption{Neutrino mixing parameters in a two-flavor oscillation scheme
where the predicted fluxes agree with the measurements in the
GALLEX/SAGE, Homestake and (Super)Kamiokande experiments.  The shaded
regions and contours are the allowed regions at 95\% C.L.~for
different axion losses ($g_{10}=0$ shaded, 4.5 solid, 10 dashed and 15
dash-dotted line). For the sake of clarity we omit the vacuum
solution for $g_{10}$=15.
\label{fig:mixing}}
\end{figure}

The additional energy sink also affects the present-day core structure
(Table~\ref{tab:runs1}) and thus the solar neutrino flux. The expected
neutrino rates are given in Table~\ref{tab:runs2} for all current
experiments. Consequently, the neutrino-oscillation parameters
{$\Delta m^2$ and $\sin^22\theta$}, for which the measurements are
reproduced, depend on $g_{10}$.  For several values of $g_{10}$ we
show in Fig.~\ref{fig:mixing} the 95\% confidence regions for the MSW
and vacuum solutions of the solar neutrino problem.
The two MSW-solutions (usually called small and large mixing angle
solutions) move toward each other with increasing axion loss. In the
case of $g_{10}$=15 there remains only one solution. 

\begin{figure}[ht]
\hbox to\hsize{\hss\epsfxsize=9.5cm\epsfbox{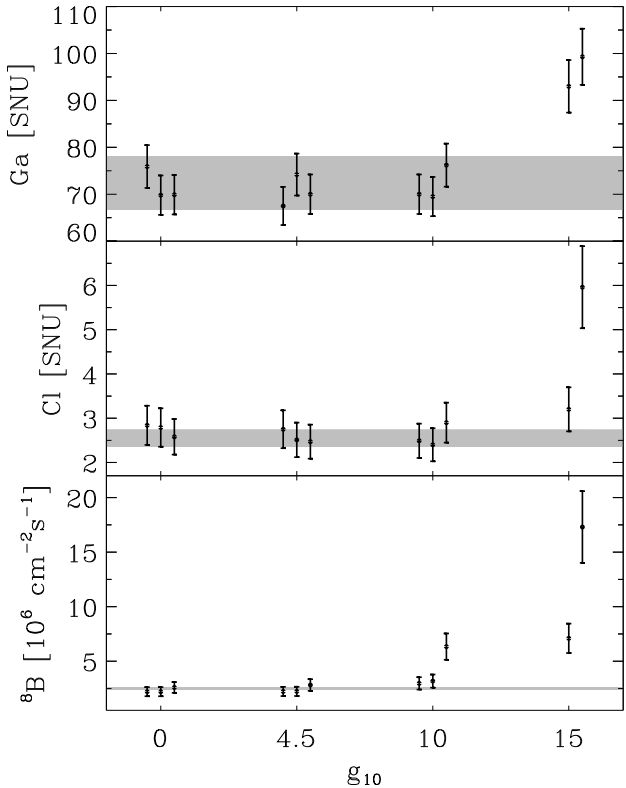}\hss}
\caption{Expected rates for GALLEX/SAGE, Homestake and
(Super)Kamiokande for different solutions and axion
losses. Theoretical uncertainties are shown by $1\sigma$-error bars.
For each $g_{10}\le10$ we plot the SMA, LMA and VAC solutions from
left to right; for $g_{10}=15$ the SMA and LMA cases have merged.  The
shaded bands represent the $1\sigma$-range of the measured rates.
\label{fig:rate}}
\end{figure}

To illustrate that for $g_{10}\geq15$ the allowed neutrino parameters
fit only very poorly the experimental values we show in
Fig.~\ref{fig:rate} the predicted event rates or neutrino fluxes for
GALLEX/SAGE, Homestake and (Super)Kamiokande. For each value of
$g_{10}$ the expected rates for the best-fit small-mixing-angle (SMA),
large-mixing-angle (LMA) and vacuum (VAC) solution are shown. The
error bars reflect the theoretical uncertainties in the predicted
rates, the shaded band the measurement errors~\cite{BP98}. In
particular, the result from (Super)Kamiokande disfavours any neutrino
oscillation solution for $g_{10}\geq15$.  Additionally, the absence of
an observed day-night effect in Super-Kamiokande excludes the region
between the SMA and LMA solution of a standard solar model
($g_{10}$=0). Including this result would lead to SMA and LMA
solutions for $g_{10}\geq10$ which reproduce the measured rates even
worse.

\section{Discussion and Summary}

We have calculated a series of self-consistent solar models with
varying amounts of axionic energy losses. For an axion-photon coupling
$g_{10}\leq4.5$, corresponding to an axion luminosity
$L_a\leq0.04\,L_\odot$, the small modification of the solar model is
well within the uncertainties of all current solar observables. On the
other hand, for $g_{10}\geq10$, corresponding to
$L_a\geq0.20\,L_\odot$, the modifications are so significant that it
is unlikely that they can be compensated by uncertainties of
standard-model input parameters such as the opacities, nuclear fusion
rates, age or metal abundance.

Our best solar model deviates from the seismic model at most 
by $\approx 0.15\%$, just within the generous uncertainties of the
seismic model. To constrain the axion-photon coupling through
helioseismology we thus only take into account the differential effect
of axion emission on solar models (Fig.~\ref{fig:uprofile}). 
For $g_{10}=10$, the sound speed deviates from
the reference model by $0.8\%$ at the local maximum of the deviation at
$r\approx0.1\,R_\odot$. This is probably too much to be absorbed
by uncertainties in the input physics of the standard solar model.

New measurements from MDI/SOHO providing more accurate low-l
frequencies should enable us to get a more precise determination
of the sound speed in the central parts. But as the difference of the
model with $g_{10}=4.5$ to the reference model ($g_{10}=0$) is very small
everywhere in the interior, these new frequencies should have almost
no influence on our limits.

Further, the $g_{10}\geq 10$ models produce less than $0.227$ for the
surface helium abundance, while the smallest value allowed by
helioseismology is 0.238. Put another way, we may no longer adjust the
presolar helium abundance to any desired value to absorb the effect of
axion losses on the solar model because the modified helium abundance
is not compatible with seismological properties of the convection
layers.

Finally, solar models with $g_{10}\geq10$ produce such large neutrino
fluxes that the measured values cannot plausibly be explained by
neutrino oscillations.  While oscillations are not yet fully
established as the undisputed explanation of the measured solar
neutrino flux deficits, it is certainly not true that any neutrino
flux can be absorbed by suitably adjusted mixing parameters.

In summary, we believe that our present knowledge of the Sun excludes
an axion-photon coupling in excess of $g_{10}=10$, corresponding to
$L_a=0.20\,L_\odot$, even if we cannot assign a statistically
meaningful confidence level to this bound.  This improves the previous
solar limit~\cite{RaffeltDearborn87} of $g_{10}\alt25$ by about a
factor of 3.  While other exotic energy-loss channels would have a
different energy and temperature dependence, it is probably generic
that helioseismology excludes any such channel much in excess of the
$0.2\,L_\odot$ level.

A recent helioscope experiment to search for solar axions has reported
a limit of $g_{10}<6$, valid for $m_a\alt0.03~{\rm eV}$
\cite{Tokyo}. Therefore, in this mass range the helioscope limit
improves our bound and thus is self-consistent, i.e.~its validity
does not depend on an axion luminosity in excess of what is allowed by
the properties of the~Sun.

Our new helioseismological bound is much weaker than the well-known
limit $g_{10}\alt0.6$ which has been derived from globular-cluster
stars~\cite{RaffeltBook,RaffeltDearborn87}.  We still think it is
useful to have independent information from the Sun, especially as the
experimental accuracy of solar observations is orders of magnitude
better than for every other star. Most importantly, the Sun serves as
a source for experimental axion searches. Note that a proposed
helioscope experiment using a decommissioned LHC test magnet could
conceivably reach the globular-cluster limit~\cite{LHC}.


\section*{Acknowledgments}

This work is based on a chapter of a thesis to be submitted by H.S.\
to the Technische Universit\"at M\"unchen in partial fulfillment of
the requirements for a doctoral degree.  Partial support by the
Deutsche For\-schungs\-ge\-mein\-schaft under grant No.\ SFB~375 is
acknowledged. We are grateful to S.~Degl'Innocenti for providing us
the uncertainties in the radial sound speed profile, Further we would
like to thank Dr.~J.N.~Bahcall for instructive comments.


\appendix

\section{Energy-Loss Rate}

\begin{figure}[ht]
\hbox to\hsize{\hss\epsfxsize=6cm\epsfbox{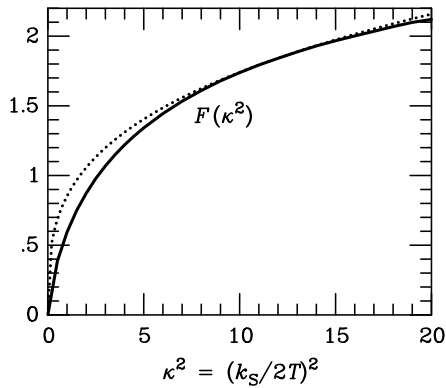}\hss}
\caption{Function $F(\kappa^2)$, solid line exact result 
 according to Eq.~(\protect\ref{eq:F}), dotted line
our approximation of Eq.~(\protect\ref{eq:Fapprox}). 
\label{fig:F}}
\end{figure}

The axionic energy-loss rate of a nondegenerate plasma by the
Primakoff effect was calculated in Ref.~\cite{Raffelt86}.  We follow
the representation given in Ref.~\cite{RaffeltBook},
\begin{eqnarray}
\epsilon&=&\frac{g_{a\gamma}^2}{4\pi}\,\frac{T^7}{\rho}\,F(\kappa^2)
\nonumber\\
&=&0.892\times10^{-3}~{\rm erg~g^{-1}~s^{-1}}\,g_{10}^2\,T_7^7\,
\rho_2^{-1}\,
F(\kappa^2),
\end{eqnarray}
where $T_7\equiv T/10^7~{\rm K}$ and $\rho_2\equiv 
\rho/10^2~{\rm g~cm^{-3}}$.
Screening effects are described by the dimensionless function
\begin{eqnarray}\label{eq:F}
F(\kappa^2)&=&\frac{\kappa^2}{2\pi^2}
\int_0^\infty dx\,\frac{x}{e^x-1}\nonumber\\
\noalign{\vskip2pt}
&&\hskip3em\times\,\left[(x^2+\kappa^2)\,\ln\left(1+
\frac{x^2}{\kappa^2}\right)-x^2\right],
\end{eqnarray}
which is shown in Fig.~\ref{fig:F}. Here,
\begin{eqnarray}
\kappa^2&=&\pi\alpha\,
\frac{n_B}{T^3}\,\Bigl(Y_e+\sum_j Z_j^2 Y_j\Bigr)
\nonumber\\
&=&16.56\,\rho_2\,T_7^{-3}\,\Bigl(Y_e+\sum_j Z_j^2 Y_j\Bigr),
\end{eqnarray}
where $\alpha=1/137$ is the fine-structure constant, $n_B$ the baryon
density, $Y_e$ the number of electrons per baryon, $Y_j$ the number of
nuclear species $j$ per baryon and $Z_j$ its charge number. In a
standard solar model $\kappa^2=12$ throughout the Sun with a variation
of less than 15\%, corresponding to $F(12)=1.842$. A good fit in this
region is given by
\begin{equation}\label{eq:Fapprox}
F(\kappa^2)=1.842\,(\kappa^2/12)^{0.31},
\end{equation}
which is shown as a dotted line in Fig.~\ref{fig:F}.

\end{document}